\input amstex
\documentstyle{amsppt}
\define\fcb{\hat\partial}
\define\inv{^{-1}}
\define\mk{\Bbb L^}
\define\sph{\Bbb S^}
\define\euc{\Bbb R^}
\define\st{\,|\,}
\define\V{\widehat V}
\define\X{\widehat X}
\redefine\L{\hat L}
\define\<{\langle}
\define\>{\rangle}
\define\LI{\text{LI}}
\define\LS{\text{LS}}
\define\An{\{A_n\}}
\define\LM{\Cal L_1(M)}
\define\BM{\Cal B(M)}
\define\BfM{\Cal B_{\text{finite}}(M)}
\define\BiM{\Cal B_\infty(M)}
\define\Bb{\partial_{\text{Bus}}}
\define\BbM{\partial_{\text{Bus}}(M)}
\define\BbfM{\partial_{\text{Bus}}^{\text{finite}}(M)}
\define\BbiM{\partial_{\text{Bus}}^{\infty}(M)}
\define\wt{\widetilde}
\define\Z{\Bbb Z}
\define\wh{\widehat}
\define\pcb{\check\partial}
\define\bcb{\bar\partial}
\define\bab{\partial_{\text{ab}}}
\define\Sz{\backsim_{\text{Sz}}}

\topmatter
\title Boundaries on Spacetimes: \\
An Outline \endtitle
\rightheadtext {Boundaries on Spacetimes}
\author
Steven G. Harris
\endauthor
\address
Department of Mathematics, Saint Louis University,
St. Louis, MO 63103, USA\endaddress
\email harrissg\@slu.edu\endemail
\abstract
The causal boundary construction of Geroch, Kronheimer,
and Penrose has some universal properties of importance
for general studies of spacetimes, particularly when
equipped with a topology derived from the causal
structure.   Properties of the causal boundary are
detailed for spacetimes with spacelike boundaries, for
multi-warped spacetimes, for static spacetimes, and for
spacetimes with group actions.  
\endabstract
\endtopmatter

\head 0. Introduction  \endhead

There is a deep history in mathematics of placing
boundaries, sometimes thought of as ideal points, on
mathematical objects which may not appear to come naturally
equipped with boundaries.  Perhaps the most famous example
is the one-point compactification of the complex
plane---i.e., the addition of a single ``point at
infinity"---resulting in the Riemann sphere.  Often, there
is more than one reasonable way to construct a boundary
for a given object, depending on the intent; for instance,
the plane---thought of as the real plane---is sometimes
equipped, not with a single point at infinity, but with a
circle at infinity, resulting in a space homeomorphic to a
closed disk.   Both these boundaries on the plane have
useful, but different, things to tell us about the nature
of the plane; the common feature is that, by bringing the
infinite reach of the plane within the confines of a more
finite object, we are better able to grasp the behavior of
the original object.

The usefulness of the construction of boundaries for an
object, in order to help realize behavior in the original
object, has not been overlooked for spacetimes.  The most
common method of constructing a suitably illuminating
boundary for a spacetime has been to embed it conformally
in a larger spacetime (often termed an unphysical
spacetime, in contrast with the original one, presumed to
have more physical meaning), and then to use the boundary
of the embedded image, with topology and causal properties
induced from the ambient (unphysical) spacetime, as the
boundary of the physical spacetime.  This is the origin,
for instance, of the usual picture of the boundary of
Minkowski space---with future-timelike and past-timelike
infinity $i^+$ and $i^-$ (each a single point),
future-null and past-null infinity $\Cal I^+$ and $\Cal
I^-$ (each a null cone), and spacelike infinity
$i^0$ (a single point)---derived from its standard
embedding into the Einstein static spacetime (see, e.g.,
\cite{HE}).

In 1972, Geroch, Penrose, and Kronheimer in \cite{GKP}
introduced a boundary-construction method for any
strongly causal spacetime, a method which was conformally
invariant, hence, a function only of the causal structure
of the spacetime, insensitive to nuances of curvature and
metric save in the grossest sense.  They called the
resulting boundary the causal boundary.  The importance of
the causal boundary seemed perhaps more to lie in the very
general nature of its construction and its apparent
naturality, rather than in any particular insights derived
from the application of this method to specific
spacetimes.  This is because the construction was rather
involved with a topology that could be quite opaque in even
simple cases.  (None the less, the causal boundary was
used for good effect in understanding the nature of the
two-dimensional trousers spacetime in \cite{HD}.)

A series of papers in recent years (\cite{Uni}, \cite{Top},
\cite{Stat}; \cite{Grp} in progress) has attempted to
demonstrate both the specific utility of the causal
boundary in a categorical sense of universality and
methods of explicating the causal boundary for wide
classes of spacetimes.  This note will summarize the
progress made to date and explore some possibilities still
being investigated.

Section 1 explores one of the motivations for believing
that the nature of the boundary of a spacetime is of
importance in understanding the global structure of a
spacetime: the possibility that invariance of spatial
topology may be related to the causal nature of the causal
boundary.

Section 2 outlines the construction of the causal boundary
and of a topology different from that considered by
Geroch, Kronheimer, and Penrose: what might be called the
future chronological topology.  The universality of the
future causal boundary construction, in terms of category
theory, is detailed.

Section 3 details the universality of the
future chronological topology for the future causal
boundary, within the limited category of spacetimes
with spacelike future boundaries.  A simple class of
examples is that of multi-warped spacetimes with spacelike
boundaries.

Section 4 shows how to construct the causal boundary for
any standard static spacetime.  These are spacetimes with,
essentially, product geometries, $\mk1 \times M$
for $M$ Riemannian.  Such spacetimes have null boundaries.

Section 5 looks at the effect on the causal boundary of
forming a quotient of a spacetime by a group action.  This
includes more general static spacetimes.

Section 6 briefly looks at some other recent work on
spacetime boundaries and mentions questions for future
investigation.

\head 1. Topology Change and Boundaries \endhead

One of the persistent questions in cosmology is whether
the spatial topology of the universe is constant in time. 
Just what is meant by ``spatial topology of the universe"
is not entirely clear, absent some very specific structure
being assumed {\it a priori\/} for the spacetime.  For 
general purposes, one can allow any suitable spacelike
hypersurface in a spacetime to be an exemplum of spatial
topology; the question then becomes, do any two
such have the same topology?

To be more specific:  ``Suitable" here probably
should mean a hypersurface which is embedded, achronal,
and edgeless in some sense.  Achronal means not just
spacelike, but that, further, no two points be
timelike-related.   ``Edgeless" can have a number of
reasonable meanings; perhaps the simplest is that the
embedding be a proper map (a sequence of points in the
domain is convergent if and only if the image sequence is
convergent in the ambient space). Then the question of
invariance of spatial topology can be formulated thus:

If a spacetime $V$ contains two spacelike hypersurfaces
$M_1$ and $M_2$, both edgeless and achronally embedded, is
it necessarily true that $M_1$ and $M_2$ are homeomorphic?

It is not hard to show that this is true for $V$ being
Minkowski space, $\mk n$, and that, in fact, any such
hypersurface must be diffeomorphic to $\euc{n-1}$; a
proof appears in \cite{Min}.  This is also true in the more
general case of a standard static spacetime $V = \mk1
\times  N$ for $N$ any Riemannian manifold; any achronal, 
edgelessly embedded spacelike hypersurface must be
diffeomorphic to $N$.  The proof of Theorem 3 in \cite{GH}
suffices for this.  (That is a far more general theorem
containing, in addition, the assumption that $V$ be
timelike or null geodesically complete, which amounts in
this case to assuming that $N$ be complete; but the proof
works without that assumption for the simplified setting of
a standard static spacetime.)

The usual boundary for Minkowski space $\mk n = \mk1
\times\euc{n-1}$---for instance, the boundary of its image
under the standard conformal embedding into the Einstein
static spacetime, $\mk1 \times \sph{n-1}$ (see
\cite{HE})---consists of a null cone in the future and a
null cone in the past (cones on the boundary sphere
$\sph{n-2}$ of $\euc{n-1}$).  For the standard static
spacetime $V = \mk1\times N$, in case $N$ is complete, the
causal boundary of $V$ is much like a null cone, a cone
on a kind of boundary at infinity on $N$, though the exact
nature of the topology is a subtle issue.  In case $N$ is
not complete, the causal boundary of $V$ is still
cone-like, but some of the cone elements are
timelike.  

Theorem 3 in \cite{GH} actually establishes invariance of
spatial topology in the much broader context of any
stationary spacetime (i.e., possessing a timelike Killing
field) which is timelike or null geodesically complete and
obeys the chronology condition (no closed timelike
curves):  Any achronal, edgelessly embedded spacelike
hypersurface must be diffeomorphic to the space of Killing
orbits.  This is generalized in \cite{HL} (Theorem 4.3,
supplemented by Theorem 2 of \cite{Mth}) to any strongly
causal spacetime $V$ possessing a foliation
$\Cal F$ by timelike curves such that every 2-sheet $S
\subset V$ ruled by $\Cal F$ has the property that,
thought of as a spacetime in its own right, every ruling
$\gamma$ in $S$ enters the past and the future of every
point of $S$:  Any achronal, edgelessly embedded spacetime
in such a $V$ must be diffeomorphic to the leafspace of
$\Cal F$. 

The causal boundary for such general spacetimes as in the
paragraph above is far from clear.  But for
a static spacetime (having a timelike
Killing field which is hypersurface-orthogonal) which is
chronological and geodesically complete, the causal
boundary is again somewhat akin to a null cone over an
appropriate object.

But it is very easy to come up with simple spacetimes
which do not exhibit invariance of spatial topology.  A
well-known example is de Sitter space, $\Bbb D^n = \{ p \in
\mk{n+1} \st p \text{ is unit-spacelike}\}$; $\Bbb D^n$ has
Cauchy surfaces which are $\sph{n-1}$ but also edgeless,
achronal, spacelike hypersurfaces which are $\euc{n-1}$.
This is actually an example (up to conformal factor) of the
more general setting of a product static spacetime $V = I
\times N$ with $I$ an interval of $\mk1$ which is finite
at one or both ends.  For instance, let $I = (-\infty,0)$. 
Then $V$ clearly has edgeless, achronal, spacelike
hypersurfaces diffeomorphic to $N$, such as $\{t\}
\times N$ for any $t<0$.  But consider any map $f: N \to
\Bbb R$ which obeys $|\text{grad}(f)| < 1$, and let $N^-
= \{x \in N \st f(x)<0\}$.  Then the map $\phi : N^- \to
V$ defined by $\phi: x \mapsto (f(x), x)$ is an edgeless,
achronal, spacelike embedding; with dimension of $N$ at
least two, we can always choose $f$ so that $N^-$ has a
different topology from that of $N$.

The boundary for de Sitter space $\Bbb D^n$, in its
conformal mapping into the Einstein static spacetime, is
a spacelike $\sph{n-1}$ for the future and the same again
for the past.  The causal boundary for $(-\infty,0)
\times N$, if $N$ is complete, is a spacelike $N$ for the
future (with something like a null cone for the past).

These examples motivate the following notion:

\proclaim{Vague Conjecture} If a spacetime has a causal
boundary which has a substantial spacelike nature, then
it does not exhibit invariance of spatial topology.  If
it has a causal boundary which is much like a null cone,
both in the future and in the past, then it does have
invariance of spatial topology.
\endproclaim

This is the sort of idea that suggests there is probably
much merit in learning the structure of the causal
boundary of as many spacetimes as possible.

\head 2. Constructions \endhead

\subhead a) Basics and Causal Structure \endsubhead

The central idea of the causal boundary of Geroch,
Kronheimer, and Penrose is to construct an endpoint for
every endless timelike curve, in such a way that the
future endpoint of a curve $\gamma$ depends only on the
curve's past $I^-[\gamma]$, and its past endpoint depends
only on its future $I^+[\gamma]$; two future-endless
timelike curves should share the same constructed future
endpoint if and only if they have the same past, and
similarly for past-endless.  (Square brackets are
employed here for a set-function defined on points to
denote its extension to sets; thus, $I^-[A]$ means
$\bigcup
\{I^-(x) \st x \in A\}$, where $I^-(x)$ is the usual past
of a point, $I^-(x) = \{y \st y \ll x\}$.  The $\ll$
relation is the usual chronology relation in a
time-oriented spacetime, $y \ll x$ meaning that there is a
timelike curve from $y$ to $x$, future-directed in that
order.) 

In the sequel, all constructions will be assumed also to
be defined in the time-dual manner, {\it
mutatis mutandis}.

The means of construction for the future causal
boundary---the future endpoints of future-endless timelike
curves---is to work with indecomposable past sets, known
as IPs.  A set $P$ is a past set if $I^-[P] = P$; it is an
indecomposable past set if it cannot be expressed as the
union of two proper past subsets.  It turns out (see, for
instance, \cite{HE}) that there are exactly two
kinds of IPs in a spacetime: the past of any point
$I^-(x)$ and the past of any future-endless timelike curve
$I^-[\gamma]$; the former are sometimes called PIPs, the
latter TIPs (for point-like and terminal IPs).  We can
then define the future causal boundary $\fcb(V)$ of a
spacetime $V$ to be, quite simply, the TIPs of $V$:
$\fcb(V) = \{P \st P \text{ is an IP and for all } x \in
V, P \neq I^-(x)\}$.

It is crucial that we get not just a set for the future
causal boundary but that there be an extension of the
chronology relation from $V$ to $\V = V \cup
\fcb(V)$ (which may be called the future completion of
$V$).  The GKP construction does this in a unified manner
for the entire future completion at once; but in many
spacetimes this introduces some new chronology relations
between the points of $V$.  This expanded notion of
chronology within $V$ itself is of importance, but it is
possible to extend $\ll$ in $V$ to $\V$ without
introducing any expansion within the spacetime proper, and
only later to make the expanded definition; and that is the
procedure followed here.

So this is the (simple) extension of $\ll$ from $V$ to
$\V$:

\roster
\item For $x \in V$ and $P \in \fcb(V)$, $x \ll P$ iff $x
\in V$.
\item For $x \in V$ and $P \in \fcb(V)$, $P \ll x$ iff $P
\subset I^-(w)$ for some $w \ll x$ ($w \in V$).
\item For $P, Q \in \fcb(V)$, $P \ll Q$ iff $P \subset
I^-(w)$ for some $w \ll x$ ($w \in V$).  
\endroster
The expanded notion of $\ll$, called here the
past-determined chronology and denoted by $\ll^p$, is
defined by including all the relations above, all pairs $x
\ll y$ in $V$, and also $x \ll^p y$ if $I^-(x) \subset
I^-(w)$ for some $w \ll y$ for $x, y, w \in V$.  Call a
spacetime past-determined if $\ll^p \;=\; \ll$\,; this
includes globally hyperbolic spacetimes and warped
products of Riemannian manifolds with past-determined
spacetimes.

The causality relation on $V$---$x \prec y$ means there is
a causal curve from $x$ to $y$, future-directed in that
order---also extends to $\V$, via $x \prec V$ iff
$I^-(x) \subset P$, $P \prec x$ iff $P \subset I^-(x)$,
and $P \prec Q$ iff $P \subset Q$.  This is not used in
any of the categorical notions.

The GKP construction continues by defining the dual
notion of the past causal boundary $\check\partial(V)$
and then detailing a very elaborate scheme for making
identifications among elements of the past and future
causal boundaries, resulting in a topology for the
combination of $V$ with the equivalence classes of
boundary points; details can be seen in \cite{HE}.  There
have been some modifications suggested for the
identification scheme, such as in \cite{BS} and, notably,
in \cite{S}.

But there are inherent problems in this combined approach,
of treating future and past causal boundary elements
together.  Very troubling, for instance, is the fact that
the topology generated for the boundary of Minkowski space
is not that of its conformal embedding into the Einstein
static spacetime:  In the GKP topology, each cone-element
(a null line) is an open set in the boundary.  Further,
there are spacetimes for which the combined
future-and-past causal boundary is neither future- nor
past-complete, in an appropriate sense.  Hence, the
approach I have followed is to deal solely with the
future causal boundary, as
$\V$ is future-complete in a strong sense, and the
future-completion construction has important universal
properties.

To see the universality, we must greatly extend the
applicability of the future completion process (details in
\cite{Uni}).  It turns out that one doesn't need much
structure to apply the completion process: only a set $X$
with a relation $\ll$ (called the chronology relation) 
such that $\ll$ is transitive and non-reflexive ($x
\not\ll x$), there are no isolates (everything is related
chronologically to something), and $X$ has a countable set
$S$ which is dense: if $x \ll y$ then for some $s \in S$,
$x \ll s \ll y$.  Call this a chronological set.  The role
of timelike curves in a chronological set is played by
future chains: sequences $\{x_n\}$ obeying $x_1 \ll \cdots
\ll x_n \ll x_{n+1} \ll \cdots$.  The role of a future
endpoint to a timelike curve is played by the notion of
future limit to a future chain $c = \{x_n\}$; this is a
point $x$ such that $I^-(x) = I^-[c]$.  This will be
unique if $X$ is past-distinguishing (i.e., $I^-(x) =
I^-(y)$ implies $x = y$), such as in a strongly causal
spacetime or its future completion.

An indecomposable past set in a chronological set $X$
can be defined exactly as in a spacetime, and $P \subset
X$ is an IP if and only if $P = I^-[c]$ for some future
chain $c$.  Then the future causal boundary of $X$ is
defined as before, $\fcb(X) = \{P \subset X \st P\text{ is
an IP and for all } x \in X, P \neq I^-(x)\}$.  The $\ll$
relation extends to $\X = X \cup \fcb(X)$ exactly as
in a spacetime, and also the past-determined expansion,
$\ll^p$.

In this broader picture, $\X$ is the same sort of
creature as $X$: a chronological set.  Furthermore, the
identification of $\X$ as ``the future completion"
of $X$ can be made explicit in this sense:  Call a
chronological set future-complete if every future chain
has a future limit (not necessarily unique); then $\X$ is always future-complete.  Why it should be called
{\it the \/} future-completion of $X$ (actually, it is
$\X$ with the past-determined expansion that
deserves this title) lies in the categorical nature of the
enterprise.  

The basic category of interest is that of
{\bf ChronologicalSet}:  The objects are chronological sets
and the morphisms are set-functions which preserve the
chronology relation and future limits:  $x \ll y$ implies
$f(x) \ll f(y)$, and $x$ is a future limit of the future
chain $c$ implies $f(x)$ is a future limit of the future
chain $f[c]$; call such a function future-continuous. 
{\bf ChronologicalSet} has various full subcategories
obtained by restricting to future-complete,
past-distinguishing, or past-determined chronological sets.

Within the subcategory of past-determined  and
past-distinguishing objects, the process of
future-completion is a functor into the future-complete
subcategory; that is to say, for any future-continuous $f:
X \to Y$ between past-determined, past-distinguishing
chronological sets, there is a future-continuous map $\hat
f: \X \to \widehat Y$, and this construction preserves
composition of functions ($\X$ and $\widehat Y$ are both
past-determined and past-distinguishing, as well as
future-complete).  Indeed,
$\hat f$ is just the ``natural" extension of $f$:  The
injections $\hat\iota_X : X \to
\X$ form a natural transformation from the identity
functor to the future-completion functor, and $\hat f
\circ \hat\iota_X = \hat\iota_Y \circ f$.  Finally, if $Y$
is already future-complete, than $\hat f : \X \to Y$
is the unique future-continuous map obeying $\hat f \circ
\hat\iota_X = f$.  

This means precisely that the future-completion functor and
the $\hat\iota$ natural transformation form a left
adjoint to the forgetful functor from future-complete,
past-determined, past-distinguishing chronological sets to
past-determined, past-distin\-guishing ones.  Then by
category theory (see, for instance, \cite{M}) we know that
this functor and natural transformation are unique, up to
natural equivalence, in providing a functorial way of
future-completing the category of past-determined,
past-distinguishing chronological sets.  In effect, $\X$ is the unique minimal future-completing object for
$X$
within the category of past-determined,
past-distinguishing chronological sets.  

The restriction to past-determined objects is annoying, as
one wishes to consider spacetimes which are not
past-determined; if there are ``holes" in a spacetime, one
still expects to be able to apply a causal boundary
construction, hoping to fill in those holes.  This is
where the expansion to $\ll^p$ comes in.  This, too, can
be expressed in a functorial, natural, and universal
manner to an appropriate subcategory of chronological sets;
we just need to restrict attention to chronological sets
which are ``past-regular": $I^-(x)$
is an IP for every point $x$.  As every spacetime is
past-regular, this is a reasonable restriction.

Performing past-determination and following with
future-completion thus yields a functorial, natural, and
universal construction for forming a future-complete (and
past-determined) chronological set from a past-regular,
past-distinguishing one (and preserving the latter
qualities).  But this is not quite the GKP construction: 
That requires first doing future-completion and then
performing the past-determination expansion.  But it
turns out that the two ways of ordering are naturally
equivalent.  Thus, the GKP future causal boundary
construction provides the minimal---categorically
unique---way of future-completing a past-regular,
past-distinguishing chronological set, retaining those
qualities.

So what about the full GKP causal boundary, compounded out
of the past and future boundaries?  For any chronological
set $X$, one can extend the chronology relation not just
to $\X = X \cup \fcb(X)$, but also to $\bar X = X \cup
\fcb(X) \cup \check\partial(X)$ (that last being the past
causal boundary).  An appropriate equivalence relation
$\cong$ on $\fcb(X) \cup \check\partial(X)$ may allow for
a chronology relation on the quotient $X^* =\, \bar
X/\!\cong$\,. This is often not past-regular, as some of
the elements of $\fcb(X)$ may have become identified.

Failure of past-regularity is not necessarily a bad
thing; when this occurs for a spacetime
$V$ (example: remove a finite timelike segment from
$\mk2$), it is likely enough that $V^*$ is the desired
completion object, and not $\V$.  It may not be
future-complete, but it is likely to have a sort of
generalized notion of future completeness; this
generalized future completeness is intended to take
note of the situation of a future chain having a
future limit in $\V$, but that future limit being
identified with other boundary elements in $V^*$.  If it
does have this generalized future completeness (not all
spacetimes do; \cite{Uni} contains an example which does
not), then upon deleting the elements of $V^*$ that have
no past, the remainder is a quotient of $\V$.  This has a
generalization in categorical terms:  The uniqueness of the
extension to $\X$ of a future-continuous map from $X$ into
a future-complete $Y$ is replicable for the appropriate
generalized notions derived from identifications.

\subhead b) Topology \endsubhead

As shown in \cite{Top}, the topology imputed to the causal
boundary of Minkowski space in \cite{GKP} does not at all
match the common expectation of a cone, such as one
obtains by taking the inherited topology from conformally
embedding Minkowski space in the Einstein static
spacetime.  Modifications of the GKP process, such as in
\cite{BS} and
\cite{S}, have no different effect in such a simple
spacetime.

The process presented here, summarizing the results of
\cite{Top}, is applicable for any past-regular,
past-distinguishing chronological set.  This topology
seems to do the right thing in a number of contexts.  It
has especially good qualities (such as universality) in
the category of chronological sets with spacelike
boundaries, as described in section 3.  This topology can
be generalized to deal with non-past-regular chronological
sets, also, but that will not be detailed here (it appears
in \cite{Top}).

The heart of the topology to be defined on a chronological
set $X$ is not any notion of open set, but of what may be
called a limit-operator on sequences in $X$:  If $\sigma =
\{x_n\}$ is a sequence of points in $X$, then $L(\sigma)$
is the set of ``first-order" limits of $\sigma$.  Under
normal circumstances, one expects $L(\sigma)$ to be either
empty or have exactly one element; but the future
chronological topology on $X$ might be non-Hausdorff (and
this is even physically reasonable, if $X = \V$ for a
spacetime $V$ in which some elements of $\fcb(V)$ ought
perhaps to be identified), in which case some sequences
will have more than one first-order limit.  In any case,
we get a topology from $L$ so long as it does the
right thing on subsequences:  For every subsequence
$\tau \subset \sigma$, we must have $L(\tau)
\supset L(\sigma)$.  Then the $L$-topology is
defined by declaring that a set $A$ is closed if and only
if for every sequence $\sigma$ in $A$, $L(\sigma)
\subset A$.

(The reason for calling the elements of $L(\sigma)$
first-order limits is that 
$L(\sigma)$ might be infinite, and for some sequence
$\tau \subset L(\sigma)$, there may be elements of
$L(\tau)$---second-order limits of $\sigma$---not appearing
in $L(\sigma)$.  For chronological sets, it takes a highly
unusual one for second-order limits to exist; in any
case, this has no effect on the development of the ideas
here.)

For the future chronological topology on a past-regular
chronological set, the limit-operator $\L$ is defined
thus:  Let $\sigma = \{x_n\}$; then $x \in \L(\sigma)$ if
and only if
\roster
\item for all $y \ll x$, for sufficiently large $n$, $y
\ll x_n$, and 
\item for any IP $P \supset I^-(x)$, if $P \neq I^-(x)$,
then for some $y \in P$, for sufficiently large $n$, $y
\not\ll x_n$.
\endroster

This can be formulated in terms of set-limits:  For any
sequence of sets $\An$, let
$$\align
\LI(\An) = \lim\inf_{n\to\infty}(\An) & =
\bigcup_{n=1}^\infty
\bigcap_{k=n}^\infty A_k \\
\LS(\An) = \lim\sup_{n\to\infty}(\An) & = 
\bigcap_{n=1}^\infty
\bigcup_{k=n}^\infty A_k
\endalign$$ 
i.e., $x \in \LI(\An)$ if and only if $x \in A_n$ for $n$
sufficiently large, and $x \in \LS(\An)$ if and only if $x
\in A_n$ for infinitely many $n$; clearly, $\LI(\An)
\subset \LS(\An)$.  If the $A_n$ are all past sets in a
chronological set, then $\LI(\An)$ and $\LS(\An)$ are past
sets, also.  Then for a sequence $\sigma = \{x_n\}$ in a
past-regular chronological set $X$, $x \in \L(\sigma)$ if
and only if 
\roster
\item $I^-(x) \subset \LI(\{I^-(x_n)\})$ and 
\item $I^-(x)$ is a maximal IP within $\LS(\{I^-(x_n)\}$. 
\endroster
In particular, if $\LI(\{I^-(x_n)\}) =
\LS(\{I^-(x_n)\})$, then in $\X$,
$\L(\sigma)$ is non-empty, as every past set contains at
least one maximal IP. \footnote{This formulation of $\hat
L$ is due to J. L. Flores.}

There are a number of important points to note about this
construction:

(1) The $\L$-topology---the future chronological
topology---is not just a topology for the
future causal boundary added to a spacetime.  Rather, it
is a topology that can be defined in any chronological set
(even past-regularity can be dispensed with); thus, a
spacetime supplemented with any sort of boundary at all,
compounded with an extension of the chronology relation to
the boundary, may be fitted with the future chronological
topology.

(2) The future chronological topology fits well with the
notion of a future limit of a future chain:  For any
future chain $c$, $\L(c)$ is precisely the set of future
limits of $c$.  It follows that in a past-distinguishing
chronological set, a future chain has at most one
topological limit in this topology.  

(3) The future chronological topology on a strongly
spacetime $V$ is precisely the same as the manifold
topology on $V$.  

(4) This topology respects (future) boundary constructions
generally:  Suppose $X$ is a past-regular chronological
set and $X$ is a subset of $\bar X$, with the chronology
relation on $X$ extending to $\bar X$, making $\bar X$
also a past-regular chronological set; further, suppose
that $X$ is chronologically dense in $\bar X$, i.e., for
all $a, b \in \bar X$ with $a \ll b$, there is some $x \in
X$ with $a \ll x \ll b$.  Then the topology induced on $X$
as a subspace of $\bar X$ (with its $\L_{\bar
X}$-topology) is the same as the $\L_X$-topology on $X$,
treated as a chronological set in its own right; and $X$ is
topologically dense in $\bar X$.

Combining points (2), (3), and (4), we see that for $V$ a
strongly causal spacetime, using the future chronological
topology for $\V$ yields the elements of
$\fcb(V)$ as topological endpoints for
future-endless timelike curves.

(5) If $V$ is a strongly causal spacetime, then in the
$\L$-topology on $\V$, $\fcb(V)$ is a closed subset.

(6) If $V = \mk n$, then the $\L$-topology on $\V$ is
precisely the same as that given by the conformal
embedding into the Einstein static spacetime, $\mk1 \times
\sph{n-1}$.  In other words:  If $\phi : V = \mk n \to E =
\mk1 \times \sph{n-1}$ is the conformal embedding, then
$\hat\phi: \V \to \widehat E$ is a homeomorphism onto its
image.  In particular, the $\L$-topology of $\fcb(\mk n)$
is that of a cone on $\sph{n-2}$.

\medskip

For a past-regular, past-distinguishing chronological set
$X$, there is not much choice when looking for a
future-complete, past-distinguishing boundary.  Although
there can be some play with respect to
past-determination, that has no effect on the
future chronological topology.  Any past-distinguishing
future completion of $X$ must be homeomorphic to $\hat X$
in the $\L$-topology.  (But non-past-distinguishing future
completions can be more desirable.)

In the generalization of future chronological topology to
non-past-regular chro\-nological sets,
points (1), (2), and (4) above remain true.  But the
strongest topological results require spacelike
boundaries. 

\head 3. Spacelike Boundaries \endhead

The categorical and universality results for {\bf
ChronologicalSet} mentioned in section 2.a are notably
absent in section 2.b.  There are, indeed, universal
results in a topological category; but we must restrict
ourselves to chronological sets with spacelike boundaries.

The problem is that if $f: X \to Y$ is a future-continuous
map between past-determined chronological sets, which is
continuous in the respective $\L$-topologies, on $X$ and
$Y$, then even though $\hat f: \X \to \widehat Y$ is
future-continuous, it is not necessarily continuous in
the $\L$-topologies on $\X$ and $\widehat Y$.  In other
words:  Even with $f$ in the topological category, $\hat
f$ may not be, thus destroying the functoriality of
future completion.  Examples of this can be found for such
simple spacetimes as $Y = \mk2$ and $X = \{(x,t) \in \mk2
\st x > 0\}$; for some continuous functions $f: X \to Y$,
the extension of $f$ to the boundary on $X$ is necessarily
discontinuous, and this can be done for $f$ preserving
$\ll$\,.  But this cannot happen when the boundary is
spacelike.

In a past-regular chronological set $X$, call a point $x$
inobservable if $I^-(x)$ is a maximal IP: no IP properly
contains it.  Then we will say that $X$ has only spacelike
boundaries (more properly: only spacelike future
boundaries) if (a) all elements of
$\fcb(X)$ are inobservable (in $\X$) and (b) $\fcb(X)$ is
closed in $\X$ or the set of inobservables of $\X$ form a
closed subset in $\X$.  (The reason for (b) is purely
technical; it is satisfied in all reasonable instances,
such as $\V$ for $V$ a strongly causal spacetime.)  A
future-continuous map $f: X
\to Y$ between past-regular, past-distinguishing
chronological sets is said to preserve spacelike
boundaries if $\hat f$ preserves inobservables.  Then the
category of interest is {\bf FutureTopology
SpacelikeBoundaries PastRegular PastDistinguishing
ChronologicalSet}: objects are past-regular,
past-distinguishing chronological sets with only
spacelike boundaries, and morphisms are future-continuous
maps that are continuous in the respective future
chronological topologies and that preserve spacelike
boundaries.  

The important result is that discontinuity can arise only
on timelike and null boundaries:   If $f$ is a morphism in
the future topological category, then so is $\hat f$. 
As all the injections $\hat\iota_X$ are in this category
when $X$ is, it follows that the categorical results from
section 2.a apply also to the topological
category above:  Future completion is a functor into the
future-complete subcategory, and future completion
together with the $\hat\iota$ injections form a left
adjoint to the forgetful functor.  

Perhaps the most interesting results lie in the category
of the generalizations for non-past-regular
chronological spaces.  All the categorical and
universality results apply, and there is also a form of
topological semi-rigidity:
  
\proclaim{Semi-rigidity of future completion}
A generalized future completion of $X$ consists of a map
$i: X \to Y$ with $i$ an isomorphism of $\ll$ onto its
image $Y_0 = i[X]$, $Y$ satisfying the generalized notion
of future-complete, and every point of $\partial(Y) = Y -
Y_0$ being a generalized future limit of a future chain
in $Y_0$.  

Suppose  $Y$ is a generalized future
completion of $X$, where $X$ and
$Y$ have only spacelike boundaries, obey the
generalization of past-distingui\-shing, and have no
points with empty pasts.  Then in the $\L$-topologies,
$Y$ is a topological quotient of $\X$.  Furthermore, if
$X$ is past-regular, then $\partial(Y)$ is a topological
quotient of $\fcb(X)$ (more generally, $\partial(Y)$ is a
topological quotient of a closely related structure). 

In particular:  If $V$ is a strongly causal spacetime
with only spacelike boundaries, then any generalized
future-completing boundary for $V$ (in other words,
anything reasonably called a sort of future completion) is
a topological quotient of $\fcb(V)$, in the
$\L$-topologies.
\endproclaim

\bigskip

A common way of providing a boundary for a spacetime is
to embed it into another spacetime and consider the
boundary of the image.  Actually, this can be done with a
topological embedding into any manifold.  If $V$ is our
spacetime and $\phi: V \to N$ is a map into a manifold
$N$ such that $\phi$ is a homeomorphism onto its image,
then for any $p \in N$, define $I^-_V(p)$ to be those
points $x \in V$ such that there is a future-directed
timelike curve from $x$ to $p$ (i.e., its $\phi$-image
approaches $p$).  Then we can consider the
$\phi$-future-boundary of $V$,
$\partial_\phi^+(V)$, to consist of those points $p \in N$
with $I^-_V(p) \neq \emptyset$.  The
$\phi$-future-completion of $V$,
$V_\phi^+$, consists of $V \cup \partial_\phi^+(V)$,
topologized by identifying $V$ with its image $\phi[V]$
in $N$.  We obtain a causal structure on $V_\phi^+$
in a manner similar to that for $\V$:   For $x \in V$ and
$p,q \in \partial_\phi^+(V)$,
 \roster
\item  $x \ll p$ if and only if $x \in I^-_V(p)$
\item $p \ll x$ if and only if for some $y \ll x$ $(y \in
V)$, $I^-_V(p) \subset I^-(y)$
\item $p \ll q$ if and only if for some $y \in I^-_V(q),$
$I^-_V(p) \subset I^-(y)$
\endroster 

This will always yield the structure of  a chronological
set for $V_\phi^+$, which therefore has a future
chronological topology, as well as the topology induced
by $\phi$.  How do these two topologies compare?  That
depends on whether $V$ has only spacelike boundaries and
on how $\phi$ extends to $\V$.

\proclaim{Embedding topology and future chronological
topology}
Let $V$ be a strongly causal spacetime with
only spacelike boundaries.  Suppose $\phi : V \to N$ is
a topological embedding of $V$ into a manifold $N$ and
that $\phi$ extends continuously to a map $\bar\phi: \V
\to N$.
\roster
\item If $\bar\phi$ is a proper map onto its image, then
the future chronological topology on $V_\phi^+$ is the
same as the $\phi$-induced topology.
\item
If the restriction of $\bar\phi$ to
$\partial_\phi^+(V)$ is a proper map onto its image, then
the future chronological topology on
$\partial_\phi^+(V)$ (from $V_\phi^+$) is the same as the
$\phi$-induced topology.
\endroster
\endproclaim

\bigskip

There is a class of spacetimes that lend themselves
to calculation of the causal boundary, so that we can
make a judgement as to whether the future chronological
topology yields something reasonable.  This is the class
of multi-warped spacetimes (or anything conformal to
such):  As a manifold, $V$ is the topological product
$(a,b) \times K_1 \times \cdots \times K_m$ for
manifolds $K_1, \dots, K_m$ ($a$ and $b$ can be
finite or infinite).  For each $i$,
$K_i$ has a Riemannian metric $h_i$ and there is a
function $f_i: (a,b) \to \euc+$.  The metric on $V$ is
$-(dt)^2 + f_1(t)h_1 + \cdots f_m(t)h_m$.  Examples (up to
conformal factor) include 
\roster
\item interior Schwarzschild: $V = (0,2m) \times \euc1
\times \sph2$ with $$(\frac{2m}r-1)(ds)^2 = -(dr)^2 +
(\frac{2m}r-1)^2(dt)^2 + r^2(\frac{2m}r-1)h_{\sph2}$$
where $h_{\sph2}$ is the usual metric on the 2-sphere
\item Robertson-Walker spacetimes: $V = (a,b) \times K$
with
$$(ds)^2 = -(dt)^2 + r(t)^2h$$
where $(K,h)$ is a quotient of $\sph3$, $\euc3$, or $\Bbb
H^3$, and $r(t)$ is the characteristic length for the
universe at time $t$
\item generalized Kasner spacetimes: $V = (0,\infty)
\times \euc1 \times \euc1 \times \euc1$ with
$$(ds)^2 = -(dt)^2 + t^{2a}(dx)^2 + t^{2b}(dy)^2 +
t^{2c}(dz)^2$$
where $a,b,c$ are constants
\endroster 

A multi-warped spacetime has only spacelike future
boundaries if and only if for each $i$, the Riemannian
metric $h_i$ is complete and the warping function grows
sufficiently large at the future end of the interval: 
Assuming that $b$ is the future end, the required
condition is that
$\int_{b^-}^b f_i^{-1/2} < \infty$ for any finite $b^- <
b$.
 
Thus, interior Schwarzschild has only spacelike future
boundaries (0 is the future end of the interval).  A
Robertson-Walker space will have only spacelike future 
boundaries if and only if
$\int_{b_-}^b 1/r(t) < \infty$, which is precisely the
condition that the spacetime be conformal to a
finite-in-the-future portion of the standard static
spacetime $\mk1 \times K$.  A generalized Kasner
spacetime has only spacelike future boundaries if and
only if the Kasner exponents
$a,b,c$ are $>1$. \footnote{This was stated incorrectly
in \cite{Uni}, but correctly given in \cite{GS}.}

\proclaim{Multi-warped spacetimes} Let $V = (a,b) \times
K_1 \times \cdots \times K_m$ be a multi-warped
spacetime, with $b$ the future end of the interval; let $K
= K_1 \times \cdots \times K_m$.  

If $V$ has only spacelike future boundaries, then in the
future chronological topology, $\V \cong (a,b]\times
K$ with $\fcb(V) \cong K$ (appearing as $\{b\} \times K$).
\endproclaim

\bigskip

Combining these three points:  Any ``reasonable"
future-completing boundary for interior
Schwarzschild is (in the future chronological topology) a
topological quotient of $\euc1\times \sph2$.  This
includes anything derived from a topological embedding
which extends to a continuous and proper map on the
future causal boundary, using the topology induced by the
embedding.

\head 4. Standard Static Spacetimes \endhead

We will consider a construction that yields the future
causal boundary for any standard static spacetime---or
anything conformal to such.  The details appear in
\cite{Stat}.

A static spacetime is one with a timelike Killing field
$U$ such that $U^\perp$ is integrable.  A standard static
spacetime is one of the form $\euc1\times M$ with metric
$g = -\Omega(dt)^2 + h$, for some positive
function $\Omega: M \to \euc+$ and some Riemannian metric
$h$ on $M$.  Because the causal boundary depends only on
the conformal class of the spacetime (being defined purely
in terms of the causal structure and topology), we
might as well confine ourselves to standard static
spacetimes with $\Omega \equiv 1$ (as $g$ above is
conformal to $-(dt)^2 + (1/\Omega)h$).  Thus, we will
look at a spacetime which is a product, $V = \mk1\times
M$ for $M$ any Riemannian manifold.

Past sets are easy to characterize in $V$.  For any
function $f: M \to \euc{}$, define $\text{P}(f) = \{(t,x)
\st t <f(x)\}$.  Then $\text{P}(f)$ is a past set if and
only if $f$ is a Lipschitz-1 function (i.e., $|f(x)-f(y)|
< d(x,y)$, where $d$ is the distance function on $M$);
and every past set arises in this manner, save for the
past set which is $V$ itself---which can be represented
as $\text{P}(\infty)$.

The IPs of $V$ are sets of the form $\text{P}(f)$ for
particularly special Lipschitz-1 functions: what might be
called the Busemann functions on $M$.  These come about
as follows:

Every IP $P$ is generated as the past of a timelike curve
$\sigma$ in $V$, which may be parametrized as $\sigma(t)
= (t,c(t))$ for some curve $c$ in $M$ satisfying $|\dot
c(t)|<1$ ($|v|$ denotes the length of a vector $v$ in $M$
using the Riemannian metric there).  Let
$[\alpha,\omega)$ be the domain of $\sigma$ in this
parametrization (with the possibility that $\omega =
\infty$, but with $\alpha$ finite), hence, the domain of
$c$.  Then a calculation yields that $P = I^-[\sigma] =
\text{P}(b_c)$, where $b_c : M \to \euc{}$ is given by
$b_c(x) = \lim_{t\to\omega}(t-d(c(t),x))$.  Actually,
$b_c$ may be infinite-valued; but either $b_c(x) =\infty$
for all $x\in M$, or $b_c$ is finite-valued on all of
$M$; and in the latter case, it is Lipschitz-1.  If $P =
I^-((s,x))$, then $\omega = s$, $x = \lim_{t\to s}c(t)$,
and $b_c = d^s_x : y \mapsto s - d(x,y)$.  The future
causal boundary of $V$ consists of all the other IPs:
$\text{P}(\infty)$ (which we can call $i^+$ in imitation
of $\mk n$) and $\text{P}(b_c)$ for finite $b_c$ which is
not any $d^s_x$, i.e., such that $c(t)$ has no limit as $t
\to \omega$. 

If $c$ is a minimizing unit-speed geodesic, then the
function $b_c$ is precisely the Busemann function for $c$,
such as is used in the construction of the boundary
sphere for a Hadamard manifold: simply connected, 
non-positive-curvature, complete Riemannian manifold (see
\cite{BGS}, for example).  Accordingly, we may call $b_c$
the Busemann function for
$c$, even for a non-geodesic curve $c$.  (We could just as
easily restrict our attention to unit-speed
$c$ here, instead of less-than-unit-speed, as the only
difference is looking at null curves instead of timelike
curves to generate IPs, and either is satisfactory for
these spacetimes.  But we cannot ignore non-geodesic
curves, as there are curves $c$ such that
$b_c \neq b_\gamma$ for any geodesic $\gamma$, and
$\text{P}(b_c)$ is part of $\fcb(V)$.)

Let $\LM$ denote the Lipschitz-1 functions on $M$,
and let $\BM$ be the finite Busemann functions
which are not any $d^s_x$, i.e., functions of the form
$b_c$ for a curve $c: [\alpha, \omega) \to M$ with no
limit-point at $\omega$.  Within $\Cal B(M)$, $b_c$ is
bounded if and only if $c$ has finite length, which is
equivalent to $\omega <\infty$ (since $|\dot c| \leq 1$);
let $\BfM$ be these Busemann functions,
$\BiM$ the rest. If $M$ is complete, then 
$\BfM$ is empty.

So we can identify $\fcb(V)$ with $\BM\cup\{i+\}$, with a
splitting of $\BM$ into ``finite" and ``infinite" parts. 
But what about topology?

We can map $V$ into $\LM$ by sending $(s,x)$
to $d^s_x$ and look for a boundary of $V$ inside $\LM$. 
The natural topology on $\LM$ is as a function
space, using the compact-open topology; the
function-space topology is quite simple when
restricted to $\LM$ as convergence in that topology is
the same as point-wise convergence of the functions. 
This mapping is a topological embedding with the
function-space topology on $\LM$, and it is tempting to
use this topology for $V\cup\BM$ inside $\LM$
(identifying $V$ with its image).   

There is a real action on $\LM$, with $a \cdot f$ (for $a
\in \euc{}$ and $f \in \LM$) defined by $a\cdot f : x
\mapsto f(x) + a$.  This action preserves $\BM$ ($a\cdot
b_c = b_{c^a}$ for $c^a(t) = c(t-a)$) and is reflected in
$V$ by $a \cdot (s,x) = (s+a, x)$.  As $V/\euc{} = M$,
one can look to $\LM/\euc{}$ for various boundaries on
$M$.  For instance, since $\LM/\euc{}$ is compact (using
the function-space topology on $\LM$), one can achieve a
compactification of $M$ by looking for its closure in
$\LM/\euc{}$; one might call the boundary of $M$ obtained
thereby the Lipschitz boundary of $M$,
$\partial_{\text{Lip}}(M)$.  Let us call $\BM/\euc{}$
the Busemann boundary of $M$, $\BbM$ (whether using the
function-space topology from $\LM$ or some other
topology).  While $\partial_{\text{Lip}}(M)$ has some
claim on us as a natural boundary on $M$, it is $\BbM$
that is central to the causal boundary on $\mk1 \times M$.

Let $\BbfM = \BfM/\euc{}$ and $\BbiM = \BiM/\euc{}$, as
those subsets of $\BM$ are also preserved
by the $\euc{}$-action.  Then $\BbfM$ represents the
Cauchy completion of $M$, while the elements of $\BbiM$
can be said to represent the points at ``geometric
infinity" for $M$, as they derive from curves which either
are rays (half-infinite geodesics minimizing along the
entire length) or behave asymptotically like rays.  If
$M$ is a Hadamard manifold, then $\BbM$ (using the
function-space topology) is precisely the boundary sphere
for $M$; but in the general setting, even for complete
$M$, $\BbM$ need not be compact or anything
like a manifold.  

Let $\pi: \LM \to \LM/\euc{}$ denote the projection to
the quotient.  For each point $\beta \in \BbM$,
$\pi\inv(\beta)$ is, of course, a line.  If $\beta \in
\BbfM$, then it is a line of timelike-related elements in
$\fcb(V)$, while if $\beta \in \BbiM$, then it is a line
of null-related elements; that is to say, for
$a > 0$, $\text{P}(b_c) \subset \text{P}(a \cdot b_c)$,
and if $b_c \in \BfM$, then $\text{P}(b_c) \ll 
\text{P}(a \cdot b_c)$.  There are no timelike relations
among the ``infinite" elements, though there may be other
null relations.  In particular, for $M$ complete, there
are only null relations within the future causal boundary.

In the function-space topology, any choice of $x_0 \in M$
yields an evaluation map $e: \LM \to \euc{}$, $e: f
\mapsto f(x_0)$, which is continuous.  This yields a
continuous cross-section $z: \LM/\euc{} \to \LM$ given by
$z : [f] \mapsto f - e(f)$.  The same cross-section works
for $\pi: \V -\{i^+\} \to M\cup\BbM$ and $\pi:
\fcb(V)-\{i^+\} \to \BbM$.   Since a fibre-bundle with
a cross-section is a trivial bundle (i.e., a product),
this means that $\V$ and $\fcb(V)$, apart from $i^+$,
are products.  Adding in $i^+$, we obtain that
$\fcb(V)$ is a cone on $\BbM$---a null cone, if $M$ is
complete (though there may be some other null relations
than those obtaining along each cone element). 

But that is with the function-space topology imputed to
$\fcb(V)$; and that topology may not be the future
chronological topology.  Functions which converge in the
function-space topology (point-wise convergence) always
converge in the future chronological topology, but there
may also be convergence in the $\L$-topology which is not
point-wise---and although $\fcb(V) -\{i^+\} \to \BbM$ and
$\V - \{i^+\} \to M\cup\BbM$ are still fibre bundles,
the $\L$-convergence can destroy the continuity of the
evaluation map $e$, allowing the fibre bundles to be
non-trivial.  \footnote{This corrects a misstatement in
\cite{Stat}.}  This is likely to happen when
$M$ has significant amounts of positive curvature going
out to infinity, so that there are geodesics that are not
minimizing.  

An example of such $M$ is the plane with the region
between $y = R$ and $y = -R$ ($R>1$) being changed to have
uniform positive curvature $1/R^2$ (think of the universal
cover of a grapefruit impaled by a stick). \footnote{Some
of the analysis of this space is due to J. L. Flores.}  In
the flat plane, all half-lines generate elements of
$\BM$, with parallel half-lines generating the same
elements of $\BbM$; thus, $\BbM$ is a circle at infinity. 
But with the roundness inserted, the geodesics $c(t) =
(t,a)$ (or $(-t,a)$) have $b_c = \infty$ for $|a|< R$;
thus, $\BbM$ consists of two arcs.  Let $b^+$ be the
Busemann function for $c^+(t) = (t,2R)$ and $b^-$ be that
for $c^-(t) = (t,-2R)$.  Then $b^+$ and $b^-$ are
intertwined in an interesting manner, and $\pi(b^+)$ and
$\pi(b^-)$ are not Hausdorff-separated in $M\cup\BbM$ if
we use the future chronological topology.  Let $\sigma$
be the sequence in $V = \mk1 \times M$ consisting of
$\{(n,Rn,0)\}$.  Then $\sigma$ has no limit in $\V$
using the function-space topology from $\LM$: $\sigma$
converges to an element of $\Cal L_1(M)$, but it is not in
$\BM$ (this is an example of $\BbM$ being smaller than
$\partial_{\text{Lip}}(M)$).  But in the future
chronological topology,
$\sigma$ converges to both
$b^+$ and $b^-$, and $\V -\{i^+\}$ is not a product over
$M\cup\BbM$.

The Busemann boundary can be complicated to work out in
detail, but its overall features are often fairly
clear.  For instance, $\Bb(\euc n) = \Bb(\Bbb H^n) =
\sph{n-1}$.  If $K$ is compact, then $\Bb(K) =
\emptyset$ and $\Bb(N\times K) = \Bb(N)$; and more
generally,
$\Bb(N_1\times N_2)$ can be expressed as a sort of
product involving $\Bb(N_1)$ and $\Bb(N_2)$.  The Busemann
boundary works well with connected sum:
$\Bb(N_1 \,\#\, N_2) = \Bb(N_1) \,\dot\cup\, \Bb(N_2)$
(disjoint union).  

\medskip 

In summary:

\proclaim{Structure of the future causal boundary for a
standard static spacetime}  For $V = \mk1 \times M$,
$\fcb(V)$ consists of $i^+$ and a set of other
elements that are organized as null lines and (if
$M$ is not complete) timelike lines, all joined to
$i^+$.  Aside from $i^+$, $\V$ has a free real action
extending the obvious one on $V$, yielding a line bundle
of $\fcb(V)-\{i^+\}$ over $\BbM$ and of $\V - \{i^+\}$
over $M\cup\BbM$.  In the function-space topology, these
bundles are trivial, yielding product structures for $\V$
and $\fcb(V)$ aside from $\{i^+\}$; hence, $\fcb(V)$ is a
cone on $\BbM$.  In the future chronological topology, that
product structure may not obtain, though one might still
consider $\fcb(V)$ to be cone-like.
\endproclaim

\head 5. Group Actions \endhead

Oftentimes, a spacetime $V$ of complicated topology is
easier to analyze by looking at its universal cover, $\wt
V$, which has the advantage of being simply connected. 
For instance, if $V$ is a static-complete spacetime
(i.e., possesses a complete, hypersurface-orthogonal,
timelike Killing field), then $\wt V$ is a standard
static spacetime, conformal to a product $\mk1 \times M$
(see, for instance, theorem 4 in \cite{GH}).  Since we
already know a lot about how to find the causal boundary
of standard static spacetimes, we may hope to use that to
get information on the boundary of the original
spacetime.  

The relation between $\wt V$ and $V$ is that there is a
group $G$ which acts on $\wt V$, and $V =
\wt V/G$; of course, $G = \pi_1(V)$, the fundamental
group of $V$.  So we are led to the question of how to
derive information on the boundary of a quotient of a
spacetime by a group action.  More generally, we can
inquire into the boundary of the quotient of a
chronological set $X$ by a group $G$ which acts 
on $X$, whose action preserves the chronology relation,
and which yields a chronological set for the quotient
$X/G$:  How is $\fcb(X/G)$ related to structures in $X$
and $\X$?  This is explored in \cite{Grp}.

First, what sort of group action on a chronological set
$X$ yields a chronological set for the quotient $X/G$? 
This is very simple:  We plainly want each group element
$g \in G$ to induce a chronological isomorphism on $X$
(i.e., $x \ll y$ implies $g \cdot x \ll g\cdot y$).  Then
$X/G$ will be a chronological set under the relation $[x]
\ll [y]$ whenever $x \ll g \cdot y$ for some $g \in G$
(with $[x]$ denoting the equivalence class of $x$), if and
only if for all $g \in G$ and $x \in X$, $x \not\ll
g\cdot x$; if this holds, call it a
chronological group action.  When the group action is
chronological, the projection
$\pi : X \to X/G$ ($\pi: x \mapsto [x]$) is
future-continuous.
 
Note that there are two possible topologies to consider
for $X/G$: the quotient topology (using the future
chronological topology on $X$) and the future
chronological topology, considering $X/G$ as a
chronological set in its own right.  Naturally, $\pi$ is
continuous with the quotient topology on $X/G$; but it
may well not be with the future chronological topology on
$X/G$.  If $X$ and $X/G$ are strongly causal spacetimes,
then we know that the two topologies on $X/G$ are the
same, because both are the manifold topology; but the
interesting question is with $X$ being the future
completion of a spacetime.

As might be expected, the $G$ action on $X$
extends to $\X$, so we may consider $\X/G$ and
$\fcb(X)/G$.  One might hope that there is some simple
relation between these objects and
$\widehat{X/G}$ and $\fcb(X/G)$, but this is generally not
the case, even for very simple examples.  As an
instructive example, consider
$X = \mk2$ and $G = \Z$, the integers, with the action $m
\cdot (t,x) = (t,x+m)$; then $X/G =
(\mk1\times\euc1)/\Z =
\mk1\times(\euc1/\Z) =
\mk1\times\sph1$, the Minkowski cylinder.  

We have $\fcb(\mk2) = \{i^+\}
\cup\{P^a_L \st a \in \euc{}\} \cup \{P^a_R \st a \in
\euc{}\}$, where $i^+ = \mk2$, $P^a_L = \{(t,x) \st t <
-x+a\}$, and $P^a_R = \{(t,x) \st t < x+a\}$.  The
$\Z$ action extends to the boundary by $m\cdot i^+
= i^+$, $m \cdot P^a_L = P^{a+m}_L$, and $m
\cdot P^a_R = P^{a-m}_R$.  The topology of $\fcb(\mk2)$
is that of a cone on $\sph0$ (the 0-sphere, two
points), i.e., a line.  The $\Z$-action is not free, and
the quotient topology on $\fcb(\mk2)/\Z$ is quite nasty: 
Each of the two null portions of the boundary is rolled up
into a circle, while the image of $i^+$ in that quotient
is a viciously non-Hausdorff point, whose only
neighborhood is all of those two circles.  If we look at
$\widehat{\mk2}/\Z$ as a chronological set and take its
future chronological topology, the imposed topology on
$\fcb(\mk2)/\Z$ isn't any better:  It's the wholly
indiscrete topology.

The boundary on $\mk2/\Z$ is easy to work out from the
material in section 4:  $\fcb(\mk1\times\sph1)$ is a null
cone on $\Bb(\sph1)$; since $\sph1$ is compact, its
Busemann boundary is empty.  Therefore, $\fcb(\mk2/\Z)$ is
just the single point $\{i^+\}$ and not
$\fcb(\mk2)/\Z$ in either topology.

On the other hand, consider the same $\Z$-action on lower
Minkowski space, ${\mk2}_- = \{(t,x) \st t < 0\}.$  We
have $\fcb({\mk2}_-) = \{P^a \st a \in \euc{}\}$, where
$P^a = \{(t,x) \st t < |x-a|\}$, and the $\Z$-action
extends to the boundary by $m \cdot P^a = P^{a+m}$. 
Thus, the quotient is a circle, $\fcb({\mk2}_-)/\Z =
\sph1$, and there is only one topology:  The quotient
topology on $\widehat{{\mk2}_-}/\Z$ is the same as its
topology as a chronological set in its own right.  And
the other way around is the same thing: ${\mk2}_-/\Z$
is the lower half of the Minkowski cylinder,
$(-\infty,0) \times \sph1$, and the material from section
3 shows its future boundary to be $\sph1$.  In this
instance, $\fcb(X/G) = \fcb(X)/G$ and $\widehat{X/G} =
\X/G$, and the two topologies, quotient and future
chronological, coincide on $\X/G$.

So if $\fcb(X/G)$ is not generally $\fcb(X)/G$, what is
it, and how do we find it?   The answer lies in
considering the invariant sets in $X$ that project to IPs
and boundary elements in $X/G$.   Define a
group-indecomposable past set (or GIP) in $X$ to be a
$G$-invariant past set which is not the union of two
proper subsets which are $G$-invariant past sets.  These
are precisely the sets of the form $\bigcup( G \cdot P)$
where $P$ is an IP (the notation here is $G \cdot x =
\{g\cdot x \st g \in G\}$ and $G \cdot A = \{G \cdot a
\st a \in A\}$); $\pi$ maps every GIP onto an IP, and the
inverse image of every IP is a GIP.  Define the
$G$-future causal boundary of $X$ to be $\fcb_G(X) = \{A
\subset X \st A \text{ is a GIP and for all } x \in X, A
\neq \bigcup (G \cdot I^-(x))\}$.  Then for any GIP $A$ in
$X$, $A \in\fcb_G(X)$ if and only if $\pi[A] \in
\fcb(X/G)$; similarly, for an IP $P$ in $X/G$, $P \in
\fcb(X/G)$ if and only if $\pi\inv[P] \in \fcb_G(X)$.  We
can define a bijection $\hat\pi^\partial : \fcb_G(X) \to
\fcb(X/G)$ via $\hat\pi^\partial: A \mapsto \pi[A]$, and
this can be a very handy way of determining the elements
of $\fcb(X/G)$.

But we wish to understand the topology of $\fcb(X/G)$ (in
terms of structures in $X$), and not just what its
elements are.  One way to do this is to realize the GIPs
of $X$ as IPs of another chronological set.  This can be
done by defining a new relation on $X$: set $x \ll_G y$
if and only if $x \ll g \cdot y$ for some $g \in G$. 
Then $X_G = (X, \ll_G)$ is a chronological set, called
the $G$-expansion of $X$.  It is a very strange
chronological set, being massively
non-past-distinguishing; but it has the nice property
that the IPs of $X_G$ are precisely the GIPs of $X$, and
$\fcb(X_G) = \fcb_G(X)$.  Furthermore, $X_G$ captures all
the right topological information (despite the fact that
$X_G$ itself is massively non-Hausdorff), in the
following sense:  Let $\pi_G : X_G \to X/G$ be the same
as $\pi$ on the set-level; this is continuous in the
respective $\L$-topologies, if $X$ is past-regular
(which implies that $X/G$ is, also).  Then
$\pi_G$ extends to
$\wh{\pi_G} : \widehat{X_G} \to \widehat{X/G}$, also
continuous.  This map takes the boundary to the boundary,
so we can consider the restriction $\wh{\pi_G}^\partial :
\fcb(X_G) \to \fcb(X/G)$; on the set-level, this is the
same as the map $\hat\pi^\partial: \fcb_G(X) \to
\fcb(X/G)$.  The pay-off is this:

\proclaim{Topology of the future causal boundary of $X/G$} 
If $X$ is a past-regular chronological set with a
chronological action from a set $G$, then $\fcb(X/G)$ can
be identified with $\fcb_G(X)$ or $\fcb(X_G)$; in the
respective future chronological topologies on
$X_G$ and $X/G$, the map $\wh{\pi_G}^\partial : \fcb(X_G)
\to \fcb(X/G)$ is a homeomorphism. 

Furthermore, the attachment of $\fcb(X/G)$ to $X/G$ is
exactly reflected in the attachment of $\fcb_G(X)$ to
$X$---interpreted as $\fcb(X_G)$ attaching to $X_G$---via
the map $\wh{\pi_G}: \widehat{X_G} \to \widehat{X/G}$, in
that a sequence $\sigma$ in $\widehat{X_G}$ converges to 
an element $A \in \fcb(X_G)$ if and only if
$\wh{\pi_G}[\sigma]$ converges to $\wh{\pi_G}(A)$ in $X/G$.
\endproclaim 

\medskip

Here is a typical application:

Let $V$ be a chronological static-complete spacetime.  In 
\cite{GH} it is shown that the space $M$ of Killing
orbits is a manifold ($\Pi: V \to M$ is a line bundle),
the universal cover $\wt V$ is conformal to the standard
static spacetime $\mk1 \times
\wt M$, where $\wt M$ is the space of Killing orbits in
$\wt V$, and the universal covering map
$\pi_V : \wt V\to V$ induces a map $\pi_M : \wt M \to M$
which is the universal covering map for $M$.  Let $G =
\pi_1(V)$, which is also $\pi_1(M)$, so that $\pi_V$ is
the projection $\wt V \to \wt V/G$, and $\pi_M$ is
the projection $\wt M \to \wt M/G$.  The $G$-action on
$\wt V$ splits into the action of $G$ on $\wt M$ and a
linear action on $\mk1$, i.e., a group homomorphism $\mu :
G \to \euc{}$; in other words, the action of $G$ on $\wt V
= \mk1 \times \wt M$ is given by $g \cdot (t,p) =
(t+\mu(g), g \cdot p)$.  The homomorphism $\mu$ can be
detected in $V$ as follows:  Pick any loop $c$ in $V$
representing the element $g \in G = \pi_1(V)$; then
$$\mu(g) = \int_c \alpha, \text{ where }\alpha =
-\frac{\<-,U\>}{|U|^2}$$ 
for $U$ the Killing field and $\<-,-\>$ the metric in $V$.
In fact, the metric is
$-|U|^2\alpha^2 + \Pi^*h$, where $h$ is a Riemannian
metric on $M$.  We have the restriction (from the
chronology condition on $V$) that for all $g \neq e$ (the
identity element) and all $p \in
\wt M$,  $|\mu(g)| < d(p, g\cdot p)$, where $d$ is the
distance function from $h$.  Another way to interpret
$\mu$ is as an element of $H^1_{\text{dR}}(V)$, the first
de Rahm cohomology group for $V$; it is a fundamental
algebraic invariant of the static spacetime.  

From section 4, we already know how to find the boundary
IPs in $\wt V = \mk1 \times \wt M$.  To find the boundary
IPs of $V$, we just need to discover the $G$-invariant
items.

For a concrete example, suppose we know that the space of
Killing orbits is a M\"obius strip crossed with
$\euc1$, i.e., $M =
\euc3/\Z$ with the action $m\cdot(x,y,z) = \mathbreak
(x+m,(-1)^my,z)$.  Then $\wt V$ is conformal to the product
$\mk1\times\euc3$, and the map $\pi_V: \wt V \to V$ is
projection by the $\Z$-action $m \cdot (t,x,y,z) = (t
+\mu m, x+m, (-1)^my, z)$, where $\mu$ is some real
number with $|\mu|<1$.  The elements of
the future causal boundary of $\wt V$, apart from
$i^+$, are all IPs of the form $P^a_\bold u = \{(t,p) \st
t < a + \<p,\bold u\>\}$, where $\bold u$ is a unit
vector.  For $\bold u =
\alpha\bold i + \beta\bold j + \gamma\bold k$, we have $m
\cdot P^a_\bold u = 
P^{a+(\mu-\alpha)m}_{\bold u'}$, where $\bold u' = \bold
u$ if $m$ is even, and $\bold u' = \bar\bold u =
\alpha\bold i - \beta\bold j + \gamma\bold k$ if $m$ is
odd.  In particular, we see that $\cup(\Z\cdot P^a_\bold
u) = i^+$ unless $\alpha = \mu$.  Thus, the boundary
GIPs, apart from $i^+$, are $\{Q^a_{\beta,\gamma} \st a\in
\euc{}, \beta^2 + \gamma^2 = 1 - \mu^2\}$, where
$Q^a_{\beta,\gamma} = P^a_\bold u \cup P^a_{\bar\bold u}$
for $\bold u = \mu\bold i + \beta\bold j + \gamma\bold
k$.  

In other words, $\fcb(V)$ is a null cone on
$\sph1/\Z_2$, where $\Z_2$ acts by reflection across the
$y$-axis.  Convergence to elements of this boundary is
found by looking for convergence in the pre-images in
the future completion of the $\Z$-expansion of
$\mk1\times \euc3$.

\medskip

So when is it that $\fcb(X/G) = \fcb(X)/G$ or $\wh{X/G} =
\X/G$?  And when is it that the quotient and future
chronological boundaries on $\X/G$ are the same?  The
answer lies with spacelike boundaries:

\proclaim{Group actions with spacelike boundaries} Suppose
$V$ is a spacetime with a group $G$ acting chronologically,
freely, and properly discontinuously such that 
\roster 
\item $V/G$ is strongly causal (which forces $V$ to be
also) and
\item $V$ and $V/G$ both have only spacelike boundaries.
\endroster
Then the quotient and future chronological
topologies are the same on $\V/G$.  If, in addition,
\roster
\item[3] $\wh V/G$ is past-distinguishing,
\endroster
then $\wh{V/G}$ is homeomorphic to
$\V/G$,  and $\fcb(V/G)$ is homeomorphic to $\fcb(V)/G$.
\endproclaim  

\medskip

Condition \therosteritem3 is unfortunate, in that it is
fairly opaque.  It amounts to saying that if the $G$-orbit
of a boundary IP $P$ covers another boundary IP $Q$, then
$Q$ must be an element of the $G$-orbit of $P$; this is not
something that is particularly easy to check.  Nor is it
at all clear that this needs to be stated as a
hypothesis, as no examples have appeared to date that
satisfy \therosteritem1 and \therosteritem2 but not
\therosteritem3; perhaps \therosteritem3 can be derived
from the others.

For multi-warped spacetimes, \therosteritem3 comes for free
(as the IPs are all very obvious):

\proclaim{Quotients of multi-warped spacetimes with
spacelike boundaries}  Suppose
$V$ is a strongly causal spacetime with only spacelike
boundaries which is covered, by means of a $G$-action, by
a multi-warped spacetime $W$, also with only spacelike
boundaries.  Then $\V \cong \wh W/G$ (in either topology,
as they are the same), and $\fcb(V) \cong \fcb(W)/G$.
\endproclaim 

\head 6. Other Work and Future Research \endhead

Other authors have done work in recent years in
exploration of boundary concepts.  One particularly
intriguing idea has been developed recently by Marolf and
Ross: a re-examination of the entire causal boundary,
combining future and past boundaries in a new manner;
this is detailed in \cite{MR}.  Although they express
their idea for strongly causal spacetimes, much of it
works just as well for chronological sets.  It can be
presented thus:

For $X$ a past- and future-regular chronological set,
define the Szabados relation (it first appeared
in \cite{S}) between the IPs and IFs of $X$ as follows: 
For $P$ an IP and $F$ an IF, $P  \Sz F$ if and only if (1)
$F \subset \bigcap \{I^+(x) \st x \in P\}$ and $F$ is a
maximal IF in that intersection, and (2) $P \subset
\bigcap \{I^-(x) \st x \in F\}$ and $P$ is a maximal IP in
that intersection.  For instance, for any $x \in X$,
$I^-(x) \Sz I^+(x)$ (and that is the only Szabados-related
pair involving $I^-(x)$ or $I^+(x)$).  Szabados used this
relation, extended for transitivity, to define an
equivalence relation on
$\fcb(X) \cup \pcb(X)$, as a modification of the GKP
procedure.  Marolf and Ross, however, have an entirely
different use for this relation (without any extension). 
They define a boundary for $X$ as 
$$\align
\bcb(X)  = & \{(P,F) \st P \in \fcb(X), F \in \pcb(X),
P \Sz F\} \; \cup \\
& \{(P,\emptyset) \st P \in \fcb(X) \text{ and for all } F
\in
\pcb(X), P \not\Sz F\} \; \cup \\
& \{(\emptyset,F) \st F \in \pcb(X) \text{ and for all } P
\in \fcb(X), P \not\Sz F\}.
\endalign$$
Then the MR completion of $X$ is $\bar X = X \cup
\bcb(X)$.  Treating any point $x \in X$ as the pair
$I^\pm(x) = (I^-(x),I^+(x))$, so that $\bar X$ can
be looked at as a set of ordered pairs, a chronology
relation on $\bar X$ is defined by $(P,F) \ll (P',F')$ if
and only if $F \cap P' \neq \emptyset$.  This makes $\bar
X$ into a chronological set, and the embedding of $X$
into $\bar X$ as $x \mapsto I^\pm(x)$ is a chronological
isomorphism onto its image:  No new chronological
relations are introduced into $X$ (as opposed to the GKP
construction, which ends with a past- and
future-determined chronological set).  

A few points of note on the chronological issues:  $X$ is
chronologically dense in $\bar X$, which is past- and
future-regular.  Unless $X$ has only spacelike future
boundaries, $\bar X$ is not necessarily
past-distinguishing, but it is always
past/future-distinguishing (a point is determined by its
past and its future together).  It is past- and
future-complete, though there may be more than one future
limit to a future chain or past limit to a past chain: 
For a future chain $c$, any $(I^-[c],F)$ with $I^-[c]
\Sz F$ (or $F = \emptyset$ if there is no such IF) is a
future limit for $c$.

Marolf and Ross define a topology for $\bar X$, bearing
some similarity in method to the future- and
past-chronological topologies.  For $V$ a
strongly causal spacetime, the MR topology on $\bar V$
induces the manifold topology on
$V$, which is topologically dense in $\bar V$; $\bcb(V)$
is closed.  A future limit for a timelike curve is a
topological endpoint (though there may be more than one
such).  Not only is $\bar V$ not necessarily Hausdorff
(which may be expected), but it might not even be
$\roman T_1$: points might not be closed.  If $V$ has only
spacelike future boundaries then we can compare the future
chronological topology on $\V$ with the MR topology by
mapping $\V$ into $\bar V$ via $P \mapsto (P,\emptyset)$
for $P \in \fcb(V)$; if $\V$ is Hausdorff---as seems
to be generally likely---this is a topological embedding
onto the image (so the same result for multi-warped
spacetimes obtains as in section 3).  For a standard
static spacetime $V = \mk1\times M$ with $M$ complete, the
same mapping works (as $P \Sz F$ is impossible); the MR
topology in this case is the same as the function-space
topology on $\V$.  Thus, in particular, for $V = \mk n$
the MR construction gives the same topology on $\bar V$ as
the embedding into the Einstein static spacetime.

\bigskip

An approach to boundaries with an eye towards
classification of boundary points by singularity-type is
the thrust of Scott and Szekeres in \cite{SS}.  This is
an examination of boundaries formed by topological
embeddings.  Starting with a spacetime (or, indeed, any
manifold) $V$, they define an envelopment of $V$ to be a
smooth topological embedding $\phi : V \to W$ into
another manifold of the same dimension, with open image. 
For an envelopment $\phi$, let $\partial_\phi(V)$ be the
boundary of $\phi(V)$ in the target space.  A boundary
set for $V$ is then any subset of such a boundary. The key
notion in \cite{SS} is that of an equivalence relation
among boundary sets for $V$:  If $B \subset
\partial_\phi(V)$ and $B' \subset
\partial_{\phi'}(V)$ are boundary sets for $V$, then $B
\simeq B'$ means that for all sequences of points
$\{p_n\}$ in $V$, $\{\phi(p_n)\}$ approaches a limit in
$B$ if and only if $\{\phi'(p_n)\}$ approaches a limit in
$B'$.  An abstract boundary point for $V$ is then defined
to be any equivalence class of boundary sets for $V$
which contains a singleton set $\{p\}$ as one of the
elements of the class.  The abstract boundary for $V$,
$\bab(V)$, is the set of all abstract
boundary points for $V$.  Thus, each element of $\bab(V)$
can be represented by a point in some
$\partial_\phi(V)$, but there are always larger boundary
sets, using different envelopments, which are equivalent
in terms of approach by points in $V$.

The main thrust of Scott and Szekeres is a schemata for
classification of abstract boundary points in terms of
being regularizable, points at infinity, or
singularities.  Being a point at infinity has to do with
being approachability by curves in $V$ of some particular
class $\Cal C(V)$, such as geodesics or curves of bounded
acceleration.  It must be a class that can be divided
into subclasses of finite and of infinite
parameter-length, independent of allowable change in
parametrization.  Different choices for $\Cal C(V)$
may yield different classifications. Those abstract
boundary points which can be approached only by the
curves in $\Cal C(V)$ of infinite parameter-length are
points at infinity; singularities are approachable by at
least one curve of finite parameter-length.

\bigskip

Garc\'ia-Parrado and Senovilla have introduced a study of
boundaries from envelopments (in the sense of
\cite{SS}) informed by another kind of equivalence
relation, which they call isocausality; this is
detailed in \cite{GS}.  They call two spacetimes $V$
and $V'$ isocausal, written $V \sim V'$, if there exist
diffeomorphisms $\phi: V \to V'$ and $\psi : V' \to V$,
both of which preserve the chronology relation.   Then a
causal extension of a spacetime $V$ is an envelopment
$\Phi: V \to W$ such that $V \sim \Phi[V]$.  They present
a classification of points in $\partial_\Phi(V)$ as
singularities and points at timelike or spacelike
infinity.  They present numerous detailed examples,
including negative-mass Schwarzschild,
Reissner-Nordstr\"om, generalized Kasner, and other
Bianchi-I spacetimes.

\bigskip

Finally, some directions for future research:

Preliminary work with J. L. Flores suggests that it is
possible to develop a time-symmetric chronological
topology, utilizing both $\hat L$ and $\check L$ (i.e.,
the limit-operators for the future and past chronological
topologies).  How useful is this?  How does it compare with
the MR topology on the MR completion of a spacetime?

It is possible to develop the Scott-Szekeres approach 
in the direction of chronological sets by introducing a
chronology relation on an appropriate subset of the
abstract boundary, such as those abstract boundary
points approachable by future-directed timelike curves
in the spacetime $V$ (i.e., extending the notion of
$\partial_\phi^+(V)$ in section 3 from a single envelopment
to abstract boundary points); call this the future abstract
boundary,
$\hat\bab(V)$.  One can then consider the future
chronological topology on
$V \cup
\hat\bab(V)$.  As the abstract boundary is a very unwieldy
set, this may make it possible to understand it a bit
better, though it's unclear what questions it would answer.

For $V$ a standard static spacetime $\mk1 \times M$, when
is it that $\V$ is a simple product (aside from $i^+$)
over the Busemann completion of $M$?  And when is the
function-space topology the same as the future
chronological topology for $\V$?  (These appear to be the
same question.)  Joint work with Flores is
currently aimed at trying to characterize Riemannian
manifolds $M$ for which this holds.

If $V$ is a chronological static-complete Riemannian
manifold, then, as shown in section 5, for $M$ the space
of Killing orbits,  $\wt V = \mk1 \times \wt M$, and there
are a $G$-action on $\wt M$ (with $G =\pi_1(V)$) and a
homomorphism $\mu: G \to \euc{}$, yielding a
$G$-action on $\wt V$ via $g \cdot (t, x) = (t + \mu(g),
g \cdot x)$, and $V = (\mk1 \times \wt M)/G$.  Does the
nature of $\fcb(V)$ depend on $\mu$, or is that
independent of $\mu$?  Under what circumstances is $\V$
(aside from $i^+$) a simple product over a completion of
$M$? 

If $V$ has a chronological group action from $G$ and both
$V$ and $V/G$ have only spacelike boundaries, is it
necessary to assume $\V/G$ is past-distinguishing in order
to have $\fcb(V/G) \cong \fcb(V)/G$, or does that come
automatically with spacelike boundaries?

Suppose $V$ is strongly causal and has a foliation
$\Cal F$ by timelike curves, such that in every 2-sheet
$S \subset V$ ruled by $\Cal F$, each ruling enters the
future and past of every point of
$S$ (treated as a spacetime in its own right).  Then can
one make any determination of $\fcb(V)$?  If $V$ is
globally hyperbolic, how closely related is $\fcb(V)$ to
the Cauchy surfaces?

And the Vague Conjecture of section 1:  Can one read
information on the invariance of spatial topology in $V$
from the nature of its future and past causal boundaries?

\end